\documentclass[12pt]{iopart}
\usepackage{graphicx}
\begin{document}

\title{Wang-Landau study of the random bond square Ising model with nearest- and next-nearest-neighbor interactions}

\author{N G Fytas, A Malakis and I Georgiou}

\address{Department of Physics, Section of Solid State
Physics, University of Athens, Panepistimiopolis, GR 15784
Zografos, Athens, Greece}

\begin{abstract}
We report results of a Wang-Landau study of the random bond square
Ising model with nearest- ($J_{nn}$) and next-nearest-neighbor
($J_{nnn}$) antiferromagnetic interactions. We consider the case
$R=J_{nn}/J_{nnn}=1$ for which the competitive nature of
interactions produces a sublattice ordering known as
superantiferromagnetism and the pure system undergoes a
second-order transition with a positive specific heat exponent
$\alpha$. For a particular disorder strength we study the effects
of bond randomness and we find that, while the critical exponents
of the correlation length $\nu$, magnetization $\beta$, and
magnetic susceptibility $\gamma$ increase when compared to the
pure model, the ratios $\beta/\nu$ and $\gamma/\nu$ remain
unchanged. Thus, the disordered system obeys weak universality and
hyperscaling similarly to other two-dimensional disordered
systems. However, the specific heat exhibits an unusually strong
saturating behavior which distinguishes the present case of
competing interactions from other two-dimensional random bond
systems studied previously.
\end{abstract}

\vspace{2pc} \noindent{\it Keywords}: Classical Monte Carlo
simulations, Classical phase transitions (Theory), Finite-size
scaling, Disordered systems (Theory) \maketitle

In the last three decades, the effect of quenched randomness to
the critical behavior of statistical models in two- (2D) and
three-dimensions (3D) has been the subject of intense studies.
First-order transitions are known to be dramatically softened
under the presence of quenched
randomness~\cite{imry79,aizenman89,hui89,chen95,chatelain01},
while continuous transitions may have their exponents altered
under random fields or random
bonds~\cite{hui89,harris74,chayes86}. There are some very useful
phenomenological arguments and some, perturbative in nature,
theoretical results, pertaining to the occurrence and nature of
phase transitions under the presence of quenched
randomness~\cite{hui89,dotsenko95,cardy96,jacobsen98,olson00,chatelain00}.
The most celebrated criterion is that suggested by
Harris~\cite{harris74}. This criterion relates directly the
persistence, under random bonds, of the non random behavior to the
specific heat exponent $\alpha_{p}$ of the corresponding pure
system. If $a_{p}$ is positive, then the disorder will be
relevant, i.e., under the effect of the disorder, the system will
reach a new critical behavior. Otherwise, if $a_{p}$ is negative,
disorder is irrelevant and the critical behavior will not change.
The value $\alpha_{p}=0$ is an inconclusive, marginal case. The 2D
Ising model falls into this category and although it is the most
studied case, it is still
controversial~\cite{dotsenko81,kinzel81,shalaev84,shankar87,ludwig90,
andreichenko90,kuhn94,kim94,reis96,ballesteros97,selke98,
mazzeo99,kim00,picco06,hadjiagapiou08}. In general and despite the
intense efforts of the last years on several different models, our
current understanding of the quenched randomness effects is rather
limited and the situation appears still unclear for both cases of
first- and second-order phase transitions.

The present Letter is the first investigation of the bond disorder
effects on an interesting 2D model with competing interactions. We
consider the square Ising model with nearest- ($J_{nn}$) and
next-nearest-neighbor ($J_{nnn}$) antiferromagnetic interactions
for a certain value of the coupling ratio $R=J_{nn}/J_{nnn}=1$.
For this value of $R$, the pure system undergoes a clear
second-order phase transition (from the superantiferromagnetic
(SAF) state to the paramagnetic state) and accurate estimates of
critical exponents have recently been
reported~\cite{malakis06,monroe07}. Since the value for the
critical exponent $\alpha_{p}$ of the specific heat of this
generalized Ising model is very close (almost identical, see the
discussion below) to that of the 2D three-state Potts model
($\alpha_{p}=1/3$) our choice of studying this case, closely
follows the motivation of similar numerical studies performed
earlier by Kim~\cite{kim96} and Picco~\cite{picco96} on the 2D
random bond three-state Potts ferromagnet. In other words, due to
the difficulties and possible crossover effects in the marginal
case of $\alpha_{p}=0$ (2D Ising model) it is desirable to study
here the critical behavior induced by disorder in a case where the
pure model has a positive specific heat exponent and according to
the Harris criterion~\cite{harris74} is expected to reach a new
critical behavior. Our results will be therefore profitably
compared to those of Kim~\cite{kim96} and Picco~\cite{picco96} for
the 2D random bond three-state Potts ferromagnet and possible
interesting differences may reflect aspects that are due to the
different microscopic interactions.

In zero field, the pure system under consideration, is governed by
the Hamiltonian:
\begin{equation}
\label{eq:1}
\mathcal{H}_{p}=J_{nn}\sum_{<i,j>}S_{i}S_{j}+J_{nnn}\sum_{(i,j)}S_{i}S_{j},
\end{equation}
where here both nearest- ($J_{nn}$) and next-nearest-neighbor
($J_{nnn}$) interactions are assumed to be positive. It is
well-known that the model develops at low temperatures SAF order
for $R=J_{nn}/J_{nnn}>0.5$~\cite{swedensen79,binder80} and by
symmetry the critical behavior associated with the SAF ordering is
the same under $J_{nn}\rightarrow -J_{nn}$. We will consider here
only the case $R=J_{nn}/J_{nnn}=1$, with $J_{nn}=J=1$. For this
case ($R=1$), the system undergoes a second-order phase
transition, in accordance with the commonly accepted scenario for
many years of a non-universal critical behavior with exponents
depending on the coupling ratio
$R$~\cite{swedensen79,binder80,landau85,tanaka92,minami94}. The
recent Wang-Landau~\cite{wang01} study of Malakis
\etal~\cite{malakis06} has refined earlier
estimates~\cite{binder80,tanaka92} for the correlation length
exponent $\nu$ and values very close to those of the 2D
three-state Potts model $\nu_{p}$(Potts)$=5/6$~\cite{wu82} were
obtained. From the finite-size scaling (FSS) of the pseudocritical
temperatures~\cite{malakis06} it was found that
$\nu_{p}$(SAF;$R=1$)$=0.8330(30)$ and the subsequent study of
Monroe and Kim~\cite{monroe07}, using the Fisher zeroes of the
partition function, yielded a quite matching estimate:
$\nu_{p}$(SAF;$R=1$)$=0.848(1)$. Furthermore, from the FSS of the
specific heat data an estimate for the ratio
$\alpha_{p}/\nu_{p}=0.412(5)$ was also found~\cite{malakis06}.
Finally, from the magnetic data and in accordance with an earlier
conjecture of Binder and Landau~\cite{binder80}, Malakis
\etal~\cite{malakis06} found additional evidence of the weak
universality scenario~\cite{suzuki74} and obtained the values
$\beta_{p}/\nu_{p}=0.125$ and $\gamma_{p}/\nu_{p}=1.75$. The
values of the above three ratios of exponents satisfy the
Rushbrook relation, assuming that $\nu_{p}=0.8292$, which is very
close to the estimate obtained from the shift behavior of the SAF
$R=1$ model, thus providing self consistency to the estimation
scheme. From these results, it is tempting to conjecture that the
SAF model with $R=1$ obeys the same thermal exponents with the 2D
three-state Potts model ($\nu_{p}=5/6=0.833\ldots$ and
$\alpha_{p}=1/3=0.333\ldots$~\cite{wu82}), but the respective
values of the magnetic critical exponents are different
($\beta_{p}/\nu_{p}=2/15=0.133\ldots$ and
$\gamma_{p}/\nu_{p}=26/15=1.733\ldots$~\cite{wu82}).

In the present study, we consider a particular type of bond
disorder, the same for both nearest- and next-nearest-neighbor
spins $i$ and $j$ according to the following bimodal distribution
\begin{equation}
\label{eq:2}
P(J_{ij})=\frac{1}{2}[\delta(J_{ij}-J_{1})+\delta(J_{ij}-J_{2})];\;\;
\frac{J_{1}+J_{2}}{2}=1;\;\;r=\frac{J_{2}}{J_{1}}=0.6.
\end{equation}
\begin{figure}[ht]
\centerline{\includegraphics*[width=12 cm]{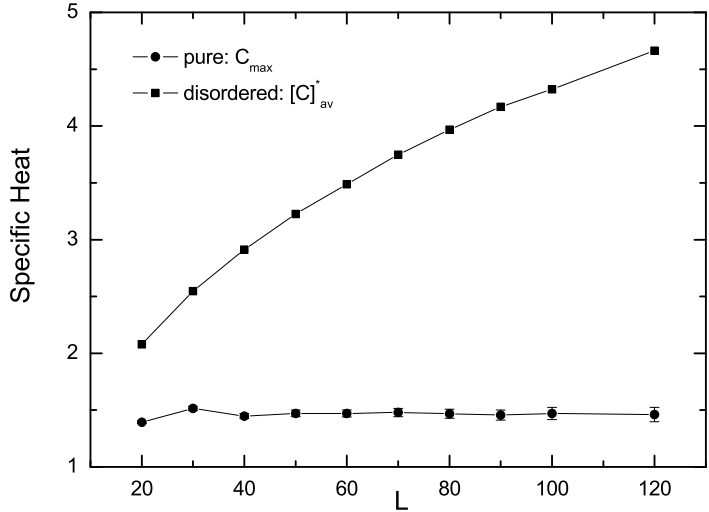}}
\caption{\label{fig:1}Size dependence of the maxima of the
specific heat for the pure (filled squares; data taken from
reference~\cite{malakis06}) and the random bond model (filled
circles).}
\end{figure}
The resulting disordered (random bond) version of the Hamiltonian
defined in equation~(\ref{eq:1}) reads now as
\begin{equation}
\label{eq:3}
\mathcal{H}=\sum_{<i,j>}J_{ij}S_{i}S_{j}+\sum_{(i,j)}J_{ij}S_{i}S_{j}.
\end{equation}
This particular choice of the disorder strength $r=0.6$ is strong
enough, as will be shown below, to observe dramatic saturation
effects on the originally diverging behavior of the specific heat
of the pure model. Only this case will be considered in this
Letter; the more general case for other values of the disorder
strength together with a comparative study with the 2D random bond
Ising model and also details of our numerical scheme will be
presented in a longer paper. The present Wang-Landau
study~\cite{wang01} closely follows our recent implementations of
an energy restricted sampling, known as critical minimum energy
subspace (CrMES)~\cite{malakis04} appropriately adapted to the
study of systems with complicated free-energy landscapes, such as
the random-field Ising model~\cite{fytas08}. We impose periodic
boundary conditions on square lattices with linear sizes $L$ in
the range $L=20-120$ and simulate relatively large ensembles of
$100$ disorder realizations.

Each disorder realization is repeatedly simulated up to $4$ times
with different initial conditions. Furthermore, in our
implementation of the Wang-Landau repetition process thermal
properties are calculated at two different Wang-Landau levels and
this practice enabled us to estimate statistical errors. The
statistical errors of the Wang-Landau method (WL-errors) used for
the estimation of thermal and magnetic properties of a particular
realization were found much smaller than the statistical errors
coming from the fact that we used, for disorder averaging, a
finite number of $100$ realizations. Therefore, the WL-errors are
not shown in our graphs, since in all cases, they are much smaller
than the symbol sizes, whereas the latter errors of ``finite
disorder sampling'' (fds-errors) are considerable and are
presented in all our figures as error bars. The mean values over
disorder are denoted as $[\ldots]_{av}$, the corresponding maxima
as $[\ldots]^{\ast}_{av}$, and the individual maxima as
$[\ldots^{\ast}]_{av}$. Since in our fitting attempts we have used
mainly data from the peaks of the disorder averaged curves (i.e.
$[C]^{\ast}_{av}$), their fds-errors are the relevant statistical
errors and have been determined by two similar methods. Using our
runs, organized in $4$ groups of $25$ realizations for each
lattice size, an application of the jackknife method~\cite{book}
and a straightforward $4$-point variance calculation (blocking
method)~\cite{book} were undertaken using the corresponding $4$
peaks of the averaged curves, for all thermal and magnetic
properties studied. It appears that the jackknife method yields
some reasonably conservative errors, that are about $15-20\%$
larger than the corresponding calculated standard deviations.
\begin{figure}[ht]
\centerline{\includegraphics*[width=16 cm]{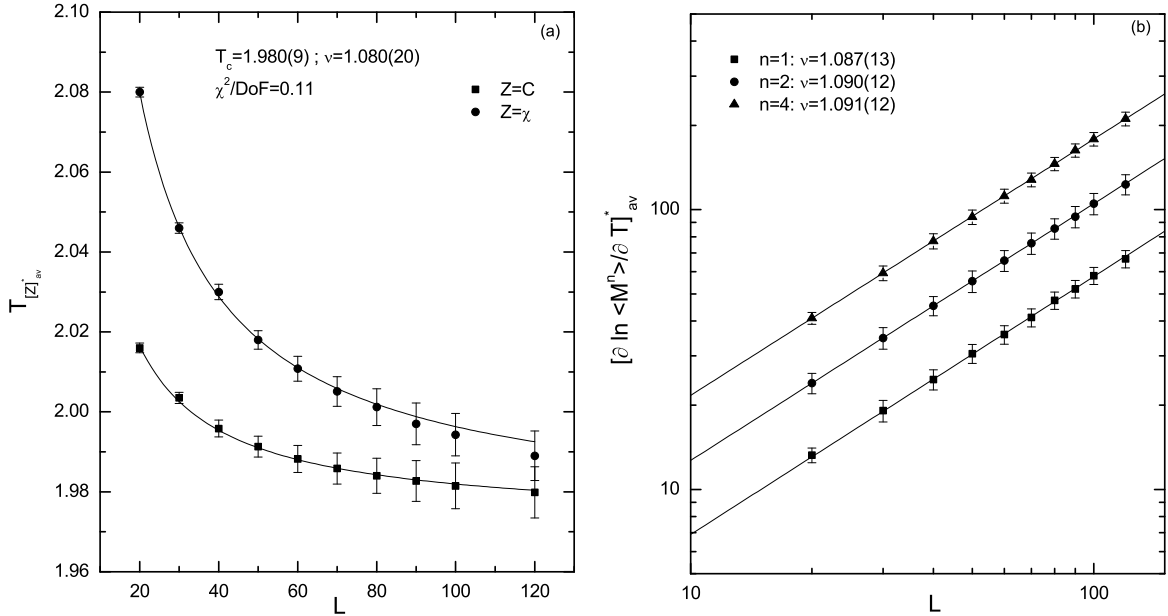}}
\caption{\label{fig:2}(a) Simultaneous fitting of the
pseudocritical temperatures of the average specific heat (filled
squares) and magnetic susceptibility (filled circles). (b) Log-log
plot of the maxima of the average logarithmic derivatives defined
in equation~(\ref{eq:5}). Linear fits are applied for $L\geq 30$.
Error bars are explained in the text.}
\end{figure}
These jackknife errors are shown as error bars in all our figures
and have been used in all our fits. Finally, let us point out that
sample-to-sample fluctuations for the individual maxima (such as
$[\chi^{\ast}]_{av}$) become very large with the lattice size. The
definition of the order-parameter follows
reference~\cite{malakis06}, using the four sublattice
magnetizations: $M=\sum_{i=1}^{4}|M_{i}|/4$.

Let us start the presentation of our results with the most
striking effect of the bond randomness on the specific heat of the
square SAF model. In figure~\ref{fig:1} we contrast the size
dependence of the specific heat maxima of the pure (filled
squares) and the random bond model (filled circles). The
suppression of the specific heat maxima is clear for the
disordered case, even for the smaller sizes shown and this
behavior should be compared with the behavior of the specific heat
of the above mentioned previous studies concerning the 2D random
bond three-state Potts ferromagnet~\cite{kim96,picco96}. It will
then be observed from these comparisons that, in full disagreement
to our finding of a strong saturating specific heat for the SAF
$R=1$ model, in the case of the 2D random bond three-state Potts
model one obtains a still diverging behavior for disorder
strengths $r=0.9$, $0.5$, and $0.25$~\cite{kim96} and an
increasing but progressively saturating behavior is obtained only
for the very strong disorder $r=0.1$~\cite{picco96}. On the other
hand, for the random bond SAF $R=1$ model, it is evident from
figure~\ref{fig:1} that the data of the average specific heat
saturate to a value already from the size of $L=40$ and any small
variation around this value is mainly coming from the fds-errors.
Therefore the estimation of the ratio $\alpha/\nu$ for this model
is not possible from the specific heat data and the alternative
route via the Rushbrook relation will be implemented later.

In figure~\ref{fig:2}(a) we present the FSS behavior of two
pseudocritical temperatures of the model $T_{[Z]^{\ast}_{av}}$,
i.e. the temperatures corresponding to the average specific heat
($Z=C$: filled squares) and the average magnetic susceptibility
($Z=\chi$: filled circles). Solid lines show a simultaneous fit on
both data according to the relation
\begin{equation}
\label{eq:4} T_{[Z]^{\ast}_{av}}=T_{c}+bL^{-1/\nu},
\end{equation}
giving $T_{c}=1.980(9)$ for the critical temperature of the
disordered model, which is to be compared with the corresponding
critical temperature $T_{c;p}=2.0823(17)$ of the pure
system~\cite{malakis06}. The value for $\chi^{2}$/DoF of the above
fit, using the jackknife errors, is $0.11$. Correspondingly, the
values of $\chi^{2}$/DoF for all our fits vary in the range
$0.1-0.4$. On the other hand, using the smaller simple standard
deviation errors, one would obtain for $\chi^{2}$/DoF values in
the range $0.2-0.7$. The above ranges for the ratios
$\chi^{2}$/DoF reflect the goodness of our fits. One could also
use the FSS of the pseudocritical temperatures defined with the
help of the average of the individual maxima of the specific heat
and susceptibility, i.e. the $T_{[Z^{\ast}]_{av}}$, but this
choice gives similar results and it is not shown here for brevity.
A first estimation of the critical exponent $\nu$ of the
correlation length is obtained from the above shift behavior and
is $\nu=1.080(20)$, as illustrated in the graph. An alternative
estimation of the exponent $\nu$ is attempted now from the FSS
analysis of the logarithmic derivatives of several powers of the
order-parameter with respect to the
temperature~\cite{chen95,ferrenberg91}
\begin{equation}
\label{eq:5} \frac{\partial \ln \langle M^{n}\rangle}{\partial
T}=\frac{\langle M^{n}E\rangle}{\langle M^{n}\rangle}-\langle
E\rangle,
\end{equation}
which scale as $L^{1/\nu}$ with the system size. In
figure~\ref{fig:2}(b) we consider in double logarithmic scale the
size dependence of the first- (filled squares), second- (filled
circles), and fourth-order (filled triangles) maxima of the
average over the ensemble of realizations logarithmic derivatives.
The solid lines shown are corresponding linear fits whose slopes
provide respectively estimates for $1/\nu$. The estimates in
figure~\ref{fig:2}(b) have an average for the correlation length
exponent of the order of $\nu=1.089$. Combining all the above
estimates we propose an error bound for $\nu$ of the order of
$0.015$. Thus, in comparison with its value of the pure model, the
exponent $\nu$ for the disordered model shows an increase of the
order of $30\%$, reflecting the strong influence of the disorder
on the thermal properties of the system. It is important to point
our here that, our estimate is in agreement with the inequality
$\nu\geq 2/D$ derived by Chayes \etal~\cite{chayes86} for
disordered systems.
\begin{figure}[ht]
\centerline{\includegraphics*[width=16 cm]{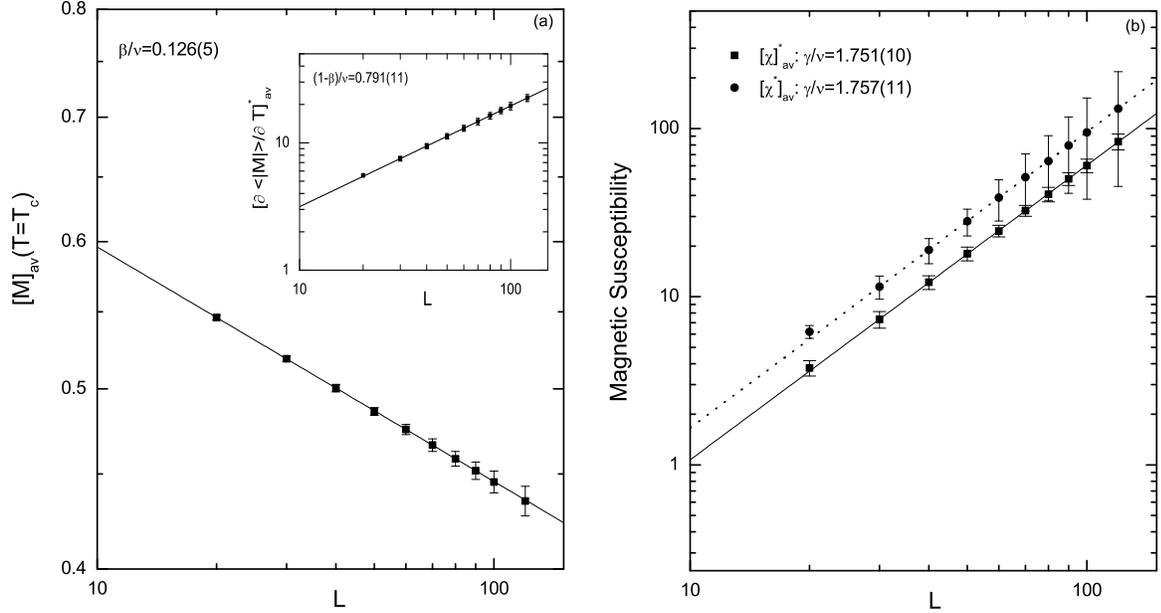}}
\caption{\label{fig:3}(a) Log-log plot of the average
magnetization at the estimated critical temperature. The inset
shows the log-log plot of the maxima of the average absolute
order-parameter derivative. (b) Log-lop plot of the size
dependence of the maxima of the magnetic susceptibilities:
$[\chi]^{\ast}_{av}$ (filled squares) and $[\chi^{\ast}]_{av}$
(filled circles). Sample-to-sample fluctuations of
$[\chi^{\ast}]_{av}$ become much larger than the fds-errors of
$[\chi]^{\ast}_{av}$.}
\end{figure}

Turning now the magnetic properties of the model we begin by
presenting the behavior of the order-parameter at the critical
temperature. We present in figure~\ref{fig:3}(a) in a log-log
scale the FSS behavior of the average order-parameter at the
estimated critical temperature $T_{c}=1.98$. The straight line
shows a linear fit for $L\geq 30$ with a slope of $0.126(5)$ which
is a first manifestation that the ratio $\beta/\nu$ has within
error bars the value of the pure model. Furthermore, in the inset
of panel (a) we plot in a log-log plot the size dependence of the
maxima of the average absolute order-parameter derivative, defined
as
\begin{equation}
\label{eq:6} \frac{\partial \langle |M|\rangle}{\partial
T}=\langle |M|E\rangle-\langle |M|\rangle\langle E\rangle,
\end{equation}
which is expected to scale as $L^{(1-\beta)/\nu}$ with the system
size~\cite{chen95,ferrenberg91}. Thus, the slope of the straight
line, which is again a linear fit for $L\geq 30$, provides the
estimate $(1-\beta)/\nu=0.791(11)$. This estimate when combined
with the value for $\nu=1.089$ gives a value for $\beta/\nu$ of
the order of $0.127$ and using the earlier error bounds for $\nu$
we propose again an error bound of the order of $0.015$. Thus, the
two estimations are self-consistent and our results indicate that
although the exponent $\beta$ increases in the disordered case,
the ratio $\beta/\nu$ remains unchanged to its pure value, i.e.
$\beta/\nu=\beta_{p}/\nu_{p}=0.125$~\cite{malakis06}. In the
sequel, we show in panel (b) of figure~\ref{fig:3} the behavior of
the magnetic susceptibility of the model in order to provide
estimates for the ratio $\gamma/\nu$. We present two data points:
the filled squares refer to the the maxima of the average curve
$[\chi]^{\ast}_{av}$, while the filled circles to the average of
the individual maxima $[\chi^{\ast}]_{av}$. In the latter case the
error bars shown reflect the sample-to-sample fluctuations, which
are, as already pointed out, larger than the fds-errors of the
corresponding averaged curves and of course much larger than the
statistical errors. The solid and dotted lines are linear fits for
$[\chi]^{\ast}_{av}$ and $[\chi^{\ast}]_{av}$ respectively, giving
the values $\gamma/\nu=1.751(10)$ and $\gamma/\nu=1.757(11)$, thus
providing convincing evidence that
$\gamma/\nu=\gamma_{p}/\nu_{p}=1.75$~\cite{malakis06}, i.e. the
ratio $\gamma/\nu$ maintains the value of the pure model. Thus,
the ratios $\beta/\nu$ and $\gamma/\nu$ for the disordered square
SAF model appear to be the same with the corresponding ratios of
the pure square SAF model but different from those of the 2D
three-state Potts model. Therefore, our results reinforce both the
weak universality scenario for the pure SAF model, as first
predicted by Binder and Landau~\cite{binder80}, as well as the
generalized statement of weak universality in the presence of bond
randomness, given by Kim~\cite{kim96} and concerning also the 2D
random bond three-state Potts ferromagnet.

Finally, as discussed above, it is not possible to directly
estimate the specific heat exponent from FSS of the specific heat
data. Yet, having estimated the values for $\beta/\nu$,
$\gamma/\nu$, and $\nu$, we may estimate $\alpha$ using either the
Rushbrook ($\alpha+2\beta+\gamma=2$) or equivalently, since
$2\beta/\nu+\gamma/\nu=2$, the hyperscaling ($2-\alpha=D \nu$)
relation. Both relations provide a negative value for the specific
heat exponent $\alpha=2-D \nu=2-2\beta-\gamma=-0.173(40)$,
reflecting the early saturation effect. This value of $\alpha$
differs significantly from the values estimated for relevant, and
even stronger, to ours randomness strength for the 2D random bond
Potts ferromagnet~\cite{kim96,picco96}. One may attribute this
strongly saturating behavior of the specific heat to the
competitive nature of interactions which is supposedly responsible
for the observed sensitivity of the SAF model to bond randomness,
since the disorder effects in this case are much more dramatic in
comparison with the effects observed in other 2D models with
simple ferromagnetic interactions, such as the 2D three-state
Potts ferromagnet~\cite{kim96,picco96}.

In conclusion, we have applied the Wang-Landau algorithm to
investigate the interesting effects caused by the presence of
quenched bond randomness on the critical behavior of the square
Ising model with nearest- and next-nearest-neighbor interactions.
Using standard finite-size scaling techniques, on high accuracy
numerical data, we have estimated the critical temperature of the
disordered model to be well below the value of the corresponding
pure model and we have extracted values for all critical exponents
of the random bond square SAF model. These values verify
hyperscaling and also satisfy the Chayes \etal
inequality~\cite{chayes86} and the weak universality scenario for
disordered systems, as stated by Kim~\cite{kim96}. The observed
unusual strong saturating behavior of the specific heat with a
negative exponent $\alpha$, distinguishes the present case of
competing interactions from other 2D random bond systems studied
previously.

\ack{Research supported by the special Account for Research Grants
of the University of Athens under Grant No. 70/4/4071. N G Fytas
acknowledges financial support by the Alexander S. Onassis Public
Benefit Foundation.}

\end{document}